\documentstyle[epsfig]{elsart}

\newcommand {\be}{\begin{equation}}
\newcommand {\ee}{\end{equation}}
\newcommand {\bx}{{\bf x}}
\newcommand {\by}{{\bf y}}
\newcommand {\bX}{{\bf X}}
\newcommand {\bY}{{\bf Y}}

\begin{document}
\bibliographystyle{unsrt}
\begin{frontmatter}
\title{A Robust Method for Detecting Interdependences:\\
  Application to Intracranially Recorded EEG}

\author[JULICH]{J. Arnhold},
\author[JULICH]{P. Grassberger},
\author[BONN]{K. Lehnertz}, 
\author[BONN]{C.E. Elger}
\address[JULICH]{John von Neumann Institute for Computing, Forschungszentrum
  J\"ulich GmbH,\\ 52425 J\"ulich, Germany}
\address[BONN]{Clinic of Epileptology, University of Bonn,
  Sigmund-Freud Str. 25, \\53105 Bonn, Germany}

\vspace{1cm}

\begin{abstract}
\baselineskip2.0em
We present a measure for characterizing statistical relationships between 
two time sequences. In contrast to commonly used measures like cross-correlations, 
coherence and mutual information, the proposed measure is non-symmetric and 
provides information about the direction of interdependence. It is 
closely related to recent attempts to detect generalized synchronization. 
However, we do not assume a strict functional relationship between the two time sequences 
and try to define the measure so as to be robust against 
noise, and to detect also weak interdependences. We apply our measure to 
intracranially recorded electroencephalograms of patients suffering from 
severe epilepsies. 

\vspace{0.5cm}
PACS numbers: 05.45.+b
\hspace{0.3cm}   87.22.-q
\hspace{0.3cm}   87.22.Jb

\vspace{0.5cm}
Keywords: nonlinear interdependence, electroencephalogram, epilepsy
\end{abstract}
\end{frontmatter}
\newpage

\section{Introduction}
\baselineskip2.0em

During the last years the analysis of synchronization phenomena received
increasing attention. Such phenomena occur in nearly all 
sciences, including physics, astrophysics, chemistry, and even economy. 
Probably the most important applications are in biology and medical sciences. 
In living systems, synchronization is often essential in normal functioning, 
while abnormal synchronization can lead to severe disorders. Typical 
examples are from neurosciences, where synchronization under normal 
conditions seems to be essential for the binding problem 
\cite{abeles82,eckhorn88,singer94,farmer98}, 
whereas epilepsies are related to abnormally strong synchronization.

Synchronization can manifest itself in different ways. At one extreme are 
coupled identical deterministic chaotic systems, which can synchronize perfectly: 
once the coupling exceeds a critical value, both systems move along identical orbits 
\cite{fujisaka83,pikovsky84}. 
If the coupled systems are not identical, in general, they can not move along 
identical orbits. If they are both chaotic and noise-free, 
a strict relationship can still exist, provided the coupling is sufficiently strong. 
Let us denote by $X=(x_1,\ldots ,x_N)$ and 
$Y=(y_1,\ldots ,y_N)$ two time sequences from which state vectors $\bx_n$ and 
$\by_n$ can be reconstructed, e.g., as delay vectors. Let us also assume that 
one of the systems, say $X$, is driving the other. By this we mean that 
the evolution of $\bx_n$ is autonomous, while $\by_{n+1}$ is a function of 
$\by_n, \bx_n$, and probably of some external noise \cite{pecora90}. If there is no noise, 
and if the driving is non-singular, $\by_{n+1} = {\bf F}(\bx_n,\by_n)$ with 
$\det(\partial F_i/\partial x_{nk})\neq 0$,  
this relationship can always be inverted (at least locally)
and can be written as $\bx_n={\bf G}(\by_n,\by_{n+1})$ or, 
after eventually increasing the embedding dimension of $Y$, as $\bx_n=\Phi(\by_n)$
\cite{schiff96}. The opposite relation (probably with some time shift $k$) 
\be
   \by_n = \Psi(\bx_{n-k})                                 \label{gen-synch}
\ee
is not guaranteed, although it looks a priori more natural in view of the 
fact that $X$ is assumed to drive $Y$. If eq.$(1)$ holds for some 
finite $k$, i.e. if the state of the driven system is a unique
function of the driver`s state, this is referred to as `generalized synchronization' 
\cite{rulkov95:_gener}. Here two cases have to be distinguished: 
strong generalized synchronization corresponds to smooth functions 
$\Psi$, while weak generalized synchronization can lead to functions which 
may even be nowhere continuous \cite{pyragas96,kocarev96}. In the latter case 
it might be difficult to detect synchronization by observing $X$ and $Y$, while 
it is immediately seen when comparing two realizations $Y^{(a)}$ and $Y^{(b)}$
of the response system:
if both are unique functions of the same $X$, then obviously $Y^{(a)}=Y^{(b)}$, 
i.e. they synchronize perfectly. 

Notice that this notion of `generalized synchronization' is closer to
the notion of interdependence, rather than to a mere time shift
generating temporal coincidences (this is what the word
synchronization actually means). If a softening of the
concept of synchronization is accepted in this way, this `generalized
synchronization' is clearly not yet the weakest and most general form
of synchronization. The weakest form is given just when $X$ and $Y$,
considered as stochastic processes, are not independent. The
problem of finding weak effects of synchronization is thus equivalent
to find weak interdependences. This is particularly true for a
system as complex as e.g., the brain, where the question wether
eq.$(1)$ holds might be meaningless.

Driver/response asymmetries, as mentioned in the above example, are indeed quite common 
also in stochastic systems. Distinguishing the driver from the responder
is of course one of the central goals, particularly in medicine 
where it is of utmost importance to detect causal relationships. Unfortunately, 
no general method exists to detect such relationships unambiguously. 
Even if $Y$ follows the motion of $X$ with a time delay as in eq.$(1)$, so that $Y$  
hardly could drive $X$, this does not proof that $X$ drives $Y$. 
Both systems might be driven by an unobserved third system $Z$. 

In particular, eq.$(1)$ by itself does not imply that $X$ drives $Y$. This is obvious 
in cases where $\Psi$ is bijective, i.e. where also $\Psi^{-1}(\cdot)$ is unique.
If $\Psi$ is not bijective (which, as we have seen, actually happens if $Y$ 
drives $X$ but fails to synchronize it), then, in general, there are several states of $X$ which 
map onto a single state of $Y$. This will typically happen if the state space of 
$X$ is larger than that of $Y$. For practical applications where strict equality 
cannot be observed but only closeness, this means that $X$ has a larger attractor 
dimension (i.e. more effective degrees of freedom) than $Y$. But this does not 
imply any causal relationship.

Typical observables used for detecting interdependences and synchronization 
are mutual information and cross correlations. Closely related to 
cross correlations are cross spectra. The main disadvantage of the latter two 
is that they measure only linear dependences. Causal relationships can (with 
the above caveats) be tested using time delays, i.e. by comparing 
$\langle x_m y_n\rangle$ with $\langle x_n y_m\rangle$. Mutual information 
is sensitive to all kinds of dependencies (it is zero only if $X$ and $Y$ 
are strictly independent), but its estimation imposes quite substantial 
requirements on the amount and quality of the data. In particular, if the 
suspected optimal embedding dimension is high, these requirements might be 
hard to meet. Finally, cross correlations and mutual information are 
symmetric in $X$ and $Y$, so that causal relationships can be detected 
{\it only} if they are associated with time delays. A priori, 
causal relationships might exist without detectable delays and, as we 
have pointed out, there might exist delays which do not reflect the naively 
expected causal relationship.

A new class of asymmetric interdependence measures which might overcome 
some of these limitations has been proposed recently 
\cite{rulkov95:_gener,schiff96,quyen98:_nonlin_eeg}. These 
authors have assumed that a deterministic relationship as in eq.$(1)$ 
exists, and have therefore not optimized their observables so as to 
detect reliably weak interdependences in a noisy environment. Moreover, 
they assumed that eq.$(1)$ automatically implies a causal relationship. That 
this is not unproblematic was discussed above. It is also seen from the 
fact that the authors of \cite{schiff96} and \cite{quyen98:_nonlin_eeg} 
drew exactly the opposite conclusions from mutual predictabilities of 
$X$ and $Y$. Equation $(1)$ was interpreted in \cite{schiff96} as 
indicating that $Y$ is the driver and $X$ the response, and that $Y$ can 
be better predicted from $X$ than vice versa. The opposite interpretation  --- 
namely that the response can be better predicted from the driver --- 
was given in \cite{quyen98:_nonlin_eeg}.
Nevertheless, these observables have been applied successfully to 
neurophysiological problems \cite{schiff96,quyen98:_nonlin_eeg}.

In the present paper we present another interdependence measure following 
closely references \cite{rulkov95:_gener,schiff96,quyen98:_nonlin_eeg}. 
But we do not assume eq.$(1)$ and we try to make our definition such as to be 
most robust. Our observable, together with several alternatives, is defined in the next section. 
Applications to EEG signals recorded from electrodes implanted under the skull 
of patients suffering from severe epilepsies are presented in Sec.3, 
while our conclusions are drawn in Sec.4.

\section{Outline of the Method}

Let $X=(x_1,x_2,\ldots,x_N)$ and $Y=(y_1,y_2,\ldots,y_N)$ denote two different 
simultaneously observed time sequences. Typically, they will be measurements of 
different observables of the same complex system, or measurements taken at 
different positions of a spatially extended system. The internal dynamics of the 
system is not known. In particular, it is not known whether the system is 
deterministic or stochastic, but we are mostly interested in cases where the 
latter is more likely a priori, or where it is at least unlikely that the 
attractor dynamics is so low that methods developed specifically for chaotic 
deterministic systems would be applicable. Physical time is related to the 
index $n$ of $x_n$, respectively $y_n$  by $t=t_0+\epsilon n$.

Time-delay embedding \cite{takens81} in an $m$-dimensional phase-space 
leads to phase-space vectors $\bx_n=(x_n,\ldots,x_{n-(m-1)\tau})$ and 
$\by_n=(y_n,\ldots,y_{n-(m-1)\tau})$. The {\it delay} $\tau$ can be chosen as 1,
but for oversampled sequences it might be useful to use some integer $\tau>1$. 
To simplify notation, we assume that also values $x_{2-m},\ldots, x_0$ and 
$y_{2-m},\ldots, y_0$ are given, so that all delay vectors with index $1\le n \le N$ 
can be formed, and the time sequences of delay vectors have $N$ elements each. The 
arrays of all delay vectors will be denoted $\bX = (\bx_1,\ldots,\bx_N)$ and 
$\bY=(\by_1,\ldots,\by_N)$.

Let $r_{n,j}$ and $s_{n,j}$, $j=1,\ldots,k$ denote the time indices of the $k$ 
nearest neighbours of $\bx_n$ and $\by_n$, respectively. Thus, the first neighbour 
distances from $\bx_n$ are $d(\bX)_n^{(1)} \equiv ||\bx_n-\bx_{r_{n,1}}|| = 
\min_q ||\bx_n-\bx_q||$, $d(\bX)_n^{(2)} \equiv ||\bx_n-\bx_{r_{n,2}}|| 
= \min_{q\neq r_{n,1}} ||\bx_n-\bx_q||$, etc., where $||\bx-\bx'||$ is the Euclidean 
distance in delay space, and similar for $\by_n$.
For each $\bx_n$, the squared mean Euclidean distance to its $k$ closest 
neighbours is defined as
\be
   R_n^{(k)}(\bX)=\frac{1}{k}\sum_{j=1}^{k}{\left( \bx_n - \bx_{r_{n,j}} \right)^2} 
                  \label{R1}
\ee 
while the {\it conditional} mean squared Euclidean distance, conditioned on the 
closest neighbour times in the time series $\bY$, is
\be
   R_n^{(k)}(\bX|\bY)=\frac{1}{k} \sum_{j=1}^{k}{\left( \bx_n - 
          \bx_{s_{n,j}} \right)^2} .                                  \label{R2}
\ee
Notice that the only difference between these two is that we used the `wrong' 
time indices for the neighbours in eq.(\ref{R2}). Instead of summing over 
nearest neighbours, we sum over those points whose equal time partners 
are nearest neighbours of $\by_n$.  Similarly we define 
\be
   R_n^{(k)}(\bY)=\frac{1}{k}\sum_{j=1}^{k}{\left( \by_n - \by_{s_{n,j}} \right)^2} 
                  \label{R1y}
\ee
and 
\be
   R_n^{(k)}(\bY|\bX)=\frac{1}{k} \sum_{j=1}^{k}{\left( \by_n - 
         \by_{r_{n,j}} \right)^2} .                                 \label{R2y}
\ee

If the point cloud $\{\bx_n\}$ has average squared radius $R(\bX) 
= \langle R^{(N-1)}(\bX)\rangle$ and effective dimension $D$ (for a 
stochastic time series embedded in $m$ dimensions, $D=m$), then 
$R_n^{(k)}(\bX)/R(\bX) \sim (k/N)^{2/D} \ll 1$ for $k\ll N$. The same is true 
for $R_n^{(k)}(\bX|\bY)$ if $\bX$ and $\bY$ are perfectly correlated, i.e. 
if there is a smooth mapping $\bx_n=\Psi(\by_n)$. On the other 
hand, if $\bX$ and $\bY$ are completely independent, then $R_n^{(k)}(\bX|\bY) 
\gg R_n^{(k)}(\bX)$. Accordingly, we introduce local and global 
interdependence measures $S_n^{(k)}(\bX|\bY)$ and $S^{(k)}(\bX|\bY)$ as 
\be
   S_n^{(k)}(\bX | \bY) \equiv \frac{R_n^{(k)}(\bX)}{R_n^{(k)}(\bX|\bY)}
                                                    \label{SXYn}
\ee
and 
\be
   S^{(k)}(\bX | \bY) \equiv \frac{1}{N} \sum_{n=1}^N S_n^{(k)}(\bX | \bY) = 
                  \frac{1}{N} \sum_{n=1}^N \frac{R_n^{(k)}(\bX)}{R_n^{(k)}(\bX|\bY)} .
                                                    \label{SXY}
\ee
Since $R_n^{(k)}(\bX|\bY)\ge R_n^{(k)}(\bX)$ by construction, we have 
\be
   0 < S^{(k)}(\bX | \bY) \le 1.
\ee
If $S^{(k)}(\bX|\bY)\approx (k/N)^{2/D} \ll 1$, then obviously $\bX$ and 
$\bY$ are independent within the limits of accuracy. If, however, 
$S^{(k)}(\bX|\bY) \gg (k/N)^{2/D}$, we say that $\bX$ depends on $\bY$,
thereby {\it without implying} any causal relationship. This dependence 
becomes maximal when $S^{(k)}(\bX|\bY)\to 1$.

The opposite dependences $S_n^{(k)}(\bY|\bX)$ and $S^{(k)}(\bY|\bX)$ are 
defined in complete analogy. They are in general {\it not} equal to 
$S_n^{(k)}(\bX|\bY)$ and $S^{(k)}(\bX|\bY)$. Both 
$S^{(k)}(\bX|\bY)$ and $S^{(k)}(\bY|\bX)$ may be of order 1. Therefore 
$\bX$ can depend on $\bY$, and at the same time can $\bY$ depend on $\bX$. 
If $S^{(k)}(\bX|\bY) > S^{(k)}(\bY|\bX)$, i.e. if $\bX$ depends 
more on $\bY$ than vice versa, we say that $\bY$ is more ``active" 
than $\bX$. Again we do {\it not} imply this to have any causal
meaning, a priori. An important question is whether an 
active/passive relationship, as defined in this way, has a causal 
driver/response interpretation in certain circumstances. 

In order to understand the origin of active/passive relationships, we 
consider the simple case where both time sequences are identical, $X=Y$, but 
we use different embedding dimensions $m_X$ and $m_Y$ in the delay vector 
construction. More precisely, we take $m_X < m_Y$ and $m_X < m_{\rm opt}$, 
where $m_{\rm opt}$ is an optimal embedding dimension in the sense
that for $m<m_{\rm opt}$ the point cloud 
$\{\bx_n\}$ is not completely unfolded, while it is unfolded for 
$m \ge m_{\rm opt}$. Thus each $\bx_n$ can be considered as a singular projection of $\by_n$, 
$\bx_n=\Psi(\by_n)$ with non-unique inverse $\Psi^{-1}$. Assume now that $\by_s$ is a close 
neighbour of $\by_n$. Then also $\bx_s$ must be a close neighbour of $\bx_n$. But the 
opposite is not true: Closeness in $\bx$ space does not imply closeness in $\by$ space.
Therefore, conditioning on times $s$ where $\by_s$ are close neighbours of $\by_n$ has less 
effect for neighbours of 
$\bx_n$ than vice versa, and $S^{(k)}(\bX|\bY) > S^{(k)}(\bY|\bX)$. Although this is not 
a mathematically rigorous argument, it shows clearly that the active/passive relationship,
as defined above, mainly reflects the relative number of degrees of freedom and not 
a driver/response relationship. Systems with 
many degrees of freedom (high dimensional ``attractors") are more active than those with few.

Notice, however, that $S^{(k)}$ is sensitive only to those degrees of freedom which are 
excited with amplitudes of order $R^{(k)}$. The latter depends, among others, on $k$ 
and on $N$. The tendency of (weakly) coupled systems to have degrees of freedom which 
are excited with very small amplitudes is well known \cite{lorenz91,torcini91}. It often leads 
to wrong estimates of attractor dimensions, and it can make the observable active/passive 
relationship to depend on parameters such as $k$ and $N$ \cite{quiroga99}. It might be responsible 
for the contradictory results of \cite{schiff96,quyen98:_nonlin,quyen98:_nonlin_eeg}.

Before leaving this section, we point out several possible generalizations and alternatives.

(a) Using the same Euclidean distance to define neighbours and in the sums in 
eqs.(\ref{R1})-(\ref{R2y}) is not necessary. Instead of the geometrical distance, 
in eqs.(\ref{R1})-(\ref{R2y})
we could have used any other dissimilarity measure between 
$\bx_n$ resp. $\by_n$ and the point clouds $\{\bx_{r_{n,j}}\}$ etc.. 
If we would have used forecasting errors in local forecasts based on these clouds, 
we would have arrived at interdependence measures very similar to those of 
\cite{schiff96,quyen98:_nonlin}. In \cite{schiff96}, also `zero time step' forecasting 
was studied. This is most closely related to our observables, but it uses only 
the distance between $\bx_n$ and the center of mass of the point cloud $\{\bx_{s_{n,j}}, 
\; j=1,\ldots k\}$, while we use all distances $|\bx_n-\bx_{s_{n,j}}|$ individually. 
It is clear that the latter contains more information, and should therefore be more 
sensitive.

(b) Instead of using arithmetic averages as in eqs.(\ref{R1})-(\ref{R2y}) and (\ref{SXY}), 
we could have used geometric or harmonic averages. And we could have replaced the average 
of ratios in eq.(\ref{SXY}) by a ratio of (arithmetic, geometric, or harmonic) averages. 
Again this could severely change sensitivity and robustness. We have not made an exhaustive 
test of all alternatives, but we checked that the above definitions are more robust than 
several alternatives. For instance, replacing eq.(\ref{SXY}) by 
\be
   S^{(k)}(\bX | \bY)' \propto \left[\frac{1}{N} \sum_{n=1}^N \frac{R_n^{(k)}(\bX|\bY)}
           {R_n^{(k)}(\bX)} \right]^{-1}
\ee
gave much more noisy results in the applications discussed in the next section which were 
also much harder to interpret physiologically. This is easily understood. In $S'$, 
occasional very small values of $R_n^{(k)}(\bX)$ have much more influence than in $S$. 
Such small values are obtained if $\bx_n$ depends abnormally weakly on $Y$, which might 
arise from some perturbation acting at time $n$. Thus $S$ is more
robust against shot noise than $S'$. We found similar results when using harmonic averages in 
eqs.(\ref{R1})-(\ref{R2y}). The main difference between the present paper and 
\cite{rulkov95:_gener} is that these authors were interested in the case of 
noiseless deterministic attractors and strong interdependences where these 
considerations play no r\^ole, and they therefore did not try do find the most 
robust observable. Also, they dicussed only the case $k=1$. This gives the strongest 
signal, but it is also much stronger affected by noise than $k>1$. In the following 
applications we used $k=10$ which seemed to give the best signal to noise ratio 
(see below).

(c) In eq.(\ref{SXYn}) we essentially compare the $\bY$-conditioned mean squared 
distances to the mean squared nearest neighbour distances. Instead of this, we 
could have compared the former to the mean squared distances to {\it random} points, 
$R_n(\bX) = (N-1)^{-1} \sum_{j\neq n} (\bx_n-\bx_j)^2$. Also, let us use the geometrical 
average in the analogon of eq.(\ref{SXY}), and define
\be
      H^{(k)}(\bX | \bY) = \frac{1}{N} \sum_{n=1}^N \log \frac{R_n(\bX)}{R_n^{(k)}(\bX|\bY)}
                                                    \label{HXY}
\ee
This is zero if $\bX$ and $\bY$ are completely independent, while it is positive if
nearness in $\bY$ implies also nearness in $\bX$ for equal time partners.
It would be negative
if close pairs in $\bY$ correspond mainly to distant pairs in $\bX$. This is
very unlikely but not impossible. Therefore, $H^{(k)}(\bX | \bY)=0$ suggests that
$\bX$ and $\bY$ are independent, but does not prove it. This (and the asymmetry under 
the exchange $\bX\leftrightarrow \bY$) is the main difference
between $H^{(k)}(\bX | \bY)$ and mutual information. The latter is strictly positive
whenever $\bX$ and $\bY$ are not completely independent. As a consequence,
mutual information is quadratic in the correlation $P(\bX,\bY)-P(\bX)P(\bY)$ for weak
correlations ($P$ are here probability distributions), while $H^{(k)}(\bX | \bY)$ is
linear. This might make $H^{(k)}(\bX | \bY)$ useful in applications.

(d) Instead of eq.(\ref{R2}) we could have defined the time shifted generalization
\be
   R_n^{(k)}(\bX|\bY,l)=\frac{1}{k} \sum_{j=1}^{k}{\left( \bx_n - \bx_{s_{n+l,j}} \right)^2} ,
                  \label{R2m}
\ee
with some (positive or negative) integer $l$.
The idea behind this definition is that it is not clear a priori that $\bx_n$ is 
most closely related to the simultaneous vector $\by_n$. Rather, if there are 
some time delays in generating either $x_n$ or $y_n$, the `natural' partner of 
$\bx_n$ might be $\by_{n+l}$. In this way we can introduce a further element of 
asymmetry which could give additional hints on causal relationships.

(e) Up to now, we have assumed in general that we use the same embedding for $\bX$ and for $\bY$. 
This is not necessary, and we could have used a different embedding dimension $m$ 
and a different delay $\tau$ for $\bY$. We did not follow this path since $\bX$ and $\bY$ had 
similar characteristics in examples studied in the next section. But it is 
worth while to point out that we can use our interdependence measure for pairs 
of time series with completely different characteristics (amplitudes, spectra, etc.). 
Dependence does not imply similarity in any sense!

(f) Instead of the Euclidean distance we could have used any other distance in defining neighbourhoods, e.g. the maximum norm.

\section{Application}
\subsection{Data Acquisition}

We analyzed electroencephalographic signals (EEG) that were recorded
in patients suffering from pharmacoresistant focal epilepsies. 
In these patients freedom of seizures can be obtained by resecting the part of the brain responsible
for seizure generation. 
Taking 
such sort of data is mandatory as part of the presurgical
analysis. The sensoring electrodes are left in the brain for typically
2 to 3 weeks. During this time the patients are 
also watched by video, so that EEG activity can be matched with behavior, and 
seizures can be identified from either. The analyses reported here were made 
after surgery had taken place, and after it had become clear from its success 
whether the localization of the epileptic focus had been correctly predicted.

EEG was recorded from electrodes implanted under 
the skull, hence close to the epileptic focus and with high 
signal-to-noise ratio. In particular, we used two types of electrodes: rectangular 
flexible grids of $8\times 8$ contacts placed onto the cortex, and pairs of needle 
shaped depth electrodes with 10 contacts each, implanted into deeper structures of 
the brain 
(see fig. 1).
\begin{figure}[ht]
\label{electrodes-scheme}
\end{figure}

EEG signals were sampled at 173 Hz using a 12 bit analog-to-digital (A/D) 
converter and filtered within a frequency band of 0.53 to 40 Hz. The cutoff frequency
of the lowpass filter was selected to suppress possible contamination by the power line.
For more details on the data and recording techniques, see \cite{lehnertz95:_spatio,lehnertz97:_neuron} and
references given therein. The data sets analyzed in this study had a duration of 10 minutes
each (cut out from much longer sequences) and were divided into segments of 
$T$ seconds each. Neighbours were searched only within the same segment.

\subsection{Parameter Selection}

As is well known, details of the delay embedding such as choice of embedding dimension 
$m$ and delay $\tau$ can be very important. In principle, the theorems of 
Takens \cite{takens81} and Sauer {\it et al.} \cite{sauer91} state that 
results should not depend on them if data are noiseless and $N$ is arbitrarily 
large, but reality tells different. Many methods have been proposed to find "optimal" parameter 
values. However, appropriate choices of $m$ and $\tau$ strongly depend on 
specific aspects of the problem at hand (such as noise level, type of 
noise, intermittency, stationarity, etc.). Thus general recipes which
do not take into account these factors can be misleading. This holds
true in particular for 
estimates of $m$ based on false nearest neighbours \cite{kennel92}. One of the most 
popular recipes \cite{fraser86:_indep} for determining the optimal delay $\tau$ 
is based on minimizing the mutual information in a {\it two}-dimensional embedding.
But in general the same $\tau$ does not minimize the mutual information in 
an embedding $m$ $\geq 3$ dimensions \cite{grassberger90}. The same comment applies 
to estimates of $\tau$ from the first zero of the autocorrelation function.
Therefore we used none of these a priori estimates of ``optimal" 
embedding parameters in this study. Instead, we approached the problem empirically by 
calculating $S^{(k)}(\bX|\bY)$ and $S^{(k)}(\bY|\bX)$ for different values 
of $m$, $\tau$, $T$, and $k$. In addition, we applied also a Theiler correction 
\cite{theiler86} by restricting the nearest neighbour times $r_{n,j}$ and 
$s_{n,j}$ to $|n-r_{n,j}| \geq \tau_{\rm Theiler}$ and 
$|n-s_{n,j}| \geq \tau_{\rm Theiler}$, and tested several values for 
$\tau_{\rm Theiler}$. It is of course not feasible to make a systematic 
search for all possible combinations of these parameters, but we feel sure 
that our final choices are reasonable and not too far from the optimum. 
We made these optimizations 
out of sample, i.e. we used a well understood `training' data set
where we could judge the reasonability of our observables by comparing with 
the medical diagnosis. This training set was not used as test set in any 
of the subsequent analyses. The ``optimal" parameters are $m=10$ (embedding dimension), $\tau=5$ (delay 
in units of sampling time), $k=10$ (neighborhood size), $T=10$ (segment length
in seconds), and $\tau_{\rm Theiler}=10$. Indeed, somewhat better results were in some 
cases obtained with larger $k$ (up to $k=100$), but we stuck to the above 
because it was faster without too much loss of significance. The delay
$\tau=5$ was implemented by simply decimating the time sequences, thereby reducing effectively the sampling rate from 173 Hz to 34.6 Hz. Thus, each 
segment contained 346 delay vectors.

\subsection{Data Representation}

\subsubsection{Depth Electrodes}

\begin{figure}[ht]
\label{S-scheme}
\end{figure}

From the 20 time sequences recorded via the depth electrodes 400 combinations 
have
to be analyzed. Results can be arranged into a $20\times 20$ interdependence
matrix $S_{ij} = S^{(k)}(\bX_i | \bX_j)$. We present our results graphically 
by means of encoding each pixel in a $20\times 20$ array using a grey scale.
Pixel $(i,j)$ is black if $S_{ij}=1$ ($\bX_i$ and $\bX_j$ are identical; this 
happens on the diagonal), while it is white if $S_{ij}=0$. The numbering of
 channels and their arrangement in the matrix are explained in fig. 2.
 
Quadrants I and IV represent interdependences between signals 
from the same (left resp. right) hemisphere, while quadrants II and III
show interdependences between different hemispheres. More precisely, 
if a pixel $(i,j)$ in quadrant II is darker than its partner $(j,i)$ in 
quadrant III, the region around contact $i$ in the right hemisphere 
is more active than the region around contact $j$ in the left hemisphere. 
Of particular interest are also average values of $S_{ij}$, i.e. averaged 
over a region symmetric under reflection along the diagonal. The average
darkness of such a region is a direct measure of its average 
interdependences with other parts of the brain
involved. 

A typical example of a grey scale pattern is shown in fig. 3 exhibiting two 
regions of high interdependence in both the 
left hemisphere and the right hemisphere. In this case the depth electrodes were not placed in a completely 
symmetrical fashion. While the electrode in the left hemisphere had 4 
contacts in the entorhinal cortex and 6 contacts in the hippocampus, 
the right electrode had 3 contacts in the entorhinal cortex and 7 
in the hippocampus. This difference (confirmed by MRI images) is clearly 
seen in fig. 3. In addition, there is a stronger interdependence
between entorhinal cortex and 
hippocampus on the left than on the right side, and the left hippocampus 
can be assumed to be more active than the right one. Interpretations  
of the latter will be given in sec.~\ref{lateralization}.

\begin{figure}[ht]
\label{matrix.fig}
\end{figure}

\subsubsection{Grid Electrodes}
\label{grid.sec}

Since grid electrodes consisted of 64 contacts, it is not very practical to 
represent the data in the same way as for the depth electrodes. In addition, 
labeling the contacts by means of a single index will result in a loss of all neighbourhood 
information, and the patterns would be hard to interpret. A different
representation is obtained by displaying each contact as a 
plaquette of an $8\times 8$ matrix, and indicating the 
activity patterns by arrows connecting these plaquettes 
\cite{quyen98:_nonlin,quyen98:_nonlin_eeg}. But also such a picture 
(which is optimal for a small number of electrodes) is too much packed 
with information for our present applications to be useful.\\
We proceeded differently. We first averaged all 
60 matrices obtained by cutting the 10 minutes recording into intervals 
of 10 seconds. The resulting time-averaged interdependences are called 
$\overline{S^{(k)}(\bX_{i_1,i_2}|\bX_{j_1,j_2})}$ where $(i_1,i_2)$ and 
$(j_1,j_2)$ are the coordinates of the contacts.
We next perform a ranking of all entries in the
$64\times 64$ matrix except the elements on the diagonal. Using the highest
one percent of entries after ranking and taking the lower end as a
cutoff $S_c$, we define for each contact $(i_1,i_2)$ an average
activity
\be
   A_{i_1,i_2} = \sum_{j_1,j_2} \overline{S^{(k)}(\bX_{j_1,j_2}|\bX_{i_1,i_2})} 
             \Theta(\overline{S^{(k)}(\bX_{j_1,j_2}|\bX_{i_1,i_2})}-S_c)
\ee
and an average passivity
\be
   P_{i_1,i_2} = \sum_{j_1,j_2} \overline{S^{(k)}(\bX_{i_1,i_2}|\bX_{j_1,j_2})} 
             \Theta(\overline{S^{(k)}(\bX_{i_1,i_2}|\bX_{j_1,j_2})}-S_c).
\ee
The cutoff $S_c$ is introduced in order to eliminate the effect of contact 
pairs with very weak interdependence. For these pairs, $\overline{S^{(k)}(\bX_i|\bX_j)}$ 
is dominated by noise, and including them would mainly decrease the 
signal-to-noise ratio. 

\begin{figure}[ht]
\label{grid.fig}
\end{figure}

Using the coordinates $i_1$ and $i_2$ we can finally represent $A_{i_1,i_2}$
and $P_{i_1,i_2}$ as $8\times 8$ grey scale matrices. 
Alternatively, we can add them and 
represent the sum $A_{i_1,i_2}+P_{i_1,i_2}$ as a grey scale matrix.
An example is given in fig. 4 exhibiting a region with very 
strong interdependence near 
the lower right corner. Its interpretation will be given in the next section.

\subsection{Results}

Our results are illustrated by three examples covering lateralization of 
the focal brain side, precise focus localization in neocortical epilepsies, 
and changes of interdependences before an impending seizure. 
These examples are quite
typical. A more systematic study involving statistically significant 
samples is under way and will be presented elsewhere.

\subsubsection{First Example}
\label{lateralization}

We analyzed 10 minutes of an interictal (seizure-free interval) EEG of a
patient suffering from a so called mesial temporal lobe epilepsy. The
clinical workup suggested the epileptic focus to be located in the left
hemisphere of the brain.
We divided the EEG data set 
into 60 nonoverlapping consecutive 10 seconds segments and 
calculated a 20 x 20 $S$-matrix for each segment as described above. 
One of these matrices was already shown in fig. 3. 
This figure is typical for all 60 matrices in showing more 
interdependences in the left hemisphere than in the right. This concerns 
both interdependences within the hippocampus, and between hippocampus 
and adjacent cortex. 
Indeed, surgery on the left side resulted in complete seizure control
of this patient. This suggests that our proposed measure 
might be able to lateralize the focal side of the brain.

\subsubsection{Second Example}

We analyzed 10 minutes of interictal EEG data from a patient suffering
from a neocortical lesional epilepsy. In this case an $8\times 8$ grid 
electrode was implanted covering the underlying brain lesion.
Again the data set was subdivided as in example one. 

A typical activity-passivity matrix obtained by means of the 
procedure described in sec.~\ref{grid.sec} is shown in fig. 4.
As already pointed out in sec.~\ref{grid.sec}, we observed highest 
interdependences in regions near the lower right corner. Indeed, 
the patient was operated on exactly in this region (which had been 
identified during presurgical evaluation) and is now free of seizures.

\begin{figure}[ht]
\label{sequence.fig}
\end{figure}

\subsubsection{Third example}

In contrast to the afore mentioned examples, where we used only EEG
recordings from a seizure free interval and averaged the data over time, 
we now study $S$ as a function of 
time. Our time resolution is again $T=10$ sec. Of particular interest are 
changes of $S$ before an impending seizure, as this could finally lead to 
its prediction \cite{lehnertz98,martinerie98:_epilep}.\footnote{The 
results of \cite{martinerie98:_epilep} use a vague definition of the 
interictal period and might therefore be questionable.}  But also changes 
during seizures and during the postictal (after-seizure) period are of interest.

A sequence of interdependence patterns taken before, during and after a seizure is shown in fig. 5. The pattern of
interdependences within the right hemisphere remains almost constant, even during 
the course of the seizure. On the other hand, $S$-values of the left hemisphere change 
dramatically. As confirmed by successful surgery, the left hemisphere was the 
focal side in this case. During the preictal stage, $S$ decreases from a high initial 
level to almost zero. Notice, that $S$ is very low also in quadrant II
directly before seizure onset, indicating that the left hemisphere is much less active. 
In frame \#13, shortly before the onset of the seizure,
interdependence builds up again on the left side. It reaches its 
maximum during the seizure and finally declines towards the interictal 
level.\\
This coincides with findings of Lehnertz and Elger \cite{lehnertz98} who
found reduced complexity before an impending seizure. Notice that 
``activity" according to our definition essentially depends on the 
number of excited degrees of freedom, which is exactly what was 
measured in \cite{lehnertz98}. The loss of activity before the 
seizure onset can be interpreted as a 
more or less hidden pathological synchronization phenomenon. 
It is assumed that seizure activity will be
induced when a "critical mass" of neurons is progressively involved in
closely time-linked high-frequency discharging. 
This critical mass might be reached if the preceeding level of synchronization 
decreases, enabling neurons to establish a synchronization which is high enough to finally lead to seizure activity. 

At first sight it may 
therefore seem paradoxical that interdependences {\it decrease} before 
a seizure. But this might indeed be exactly what happens. In a 
healthy brain a critical mass is never reached because neurons 
are strongly tied into networks where they communicate with others. 
A critical stage may be reached when a large population is 
``idle" and therefore on the one hand uncorrelated with the rest 
of the network, but on the other hand, easily recruitable for 
subsequent coherent pathophysiological activity.

\section{Discussion}

We have presented an observable which can detect dependences 
between simultaneously measured time sequences. It is 
similar to other synchronization measures proposed recently, but 
is somewhat simpler and more robust. With the other 
measures it shares the property of being asymmetric. In principle, it can be assumed
that our measure can indicate causal relationships. This might be useful
identifying the driver of the two subsystems emitting the sequences. 
We claim that such information might be 
obtainable in principle, but the interpretation is subtle and 
naive arguments can be quite misleading.
Nevertheless, this asymmetry is very interesting. 
It mainly depends on the difference in `activity' which measures 
the {\it effective} number of excited degrees of freedom. 
This {\it effective} number of active degrees of freedom depends on
the scales to which 
the observable is most sensitive. In principle, in an asymmetric 
driver-response pair the attractor dimension of the response 
is always at least as high as that of the driver (if both are 
deterministic), but this might be relevant only at length scales 
which are too small to be resolved practically.

Our measure could also be used to detect generalized synchronization, 
but we do not assume in our applications that the signals are chaotic 
with low dimensions. In contrast to recent attempts to detect 
{\it phase} synchronization in brain signals, our measure does 
not treat phase information different from amplitude information, 
and thus we cannot discuss phase or frequency locking. 

We applied our measure to intracranial multichannel EEG recordings 
taken from patients suffering 
from severe epilepsies. We found significant dependences between 
different recording sites, and these dependences were in general not 
symmetric. Due to the careful pre-operational screening of these patients 
and their observation after being operated, we could compare 
our results in detail with other neurophysiological findings. 
The most interesting preliminary results are the following: 

1) During seizure-free intervals, the seizure generating area of the brain exhibited higher interdependences than other brain areas. 

2) Some seizures analyzed here were preceeded by short periods 
(30 s to several minutes) during which extremely low dependences were
confined to the seizure generating area. 

Although these results are very encouraging, a more systematic 
study is needed and is under way. In addition, a 
host of further investigations is imaginable. Obvious candidates are the 
influences of drugs, the effect of mental activity (epilepsy patients behave 
normal even with implanted electrodes), or of various stimuli. Another important problem is the 
determination of the `critical mass´ of neurons needed 
to trigger a seizure. Moreover, a more systematic 
comparison with other diagnostical tools is necessary beforehand. Finally, the 
present findings already suggest a number of physiological results whose 
interpretation demands a thorough theoretical study. For instance, 
it is a priori not clear whether a seizure is primarily 
triggered by a change of activity in the seizure generating area, or a 
change of susceptibility of the surrounding regions. We hope 
that the near future will show progress along these lines.

\vspace{0.3cm}
{\bf Acknowledgements}

We thank J. M\"uller-Gerking, R. Quian Quiroga, T. Schreiber, and
W. Burr for the inspiring discussions and helpful comments during the study.

\bibliography{arnhold}

\newpage
{\bf Figure captions:}
\vspace{1cm}

Fig. 1: Schematic view of the two types of intracranial electrodes used in this paper. 
Grids were placed onto the cortex and have either $8\times 8$ 
electrodes. Needle shaped depth electrodes have ten
contacts each and were always used pairwise in a left-right
symmetrical fashion. In some cases, depth electrodes and grids were
used together.

\vspace{1cm}

Fig. 2: Scheme of subdivision of the 20x20 matrix $S_{ij}$. The indices 
$L_1$ to $L_{10}$ denote the contacts on the left depth electrode, 
from innermost ($L_1$) to outermost ($L_{10}$). Similarly, $R_1$ to $R_{10}$ correspond
to the right depth electrode. The index $i$ runs horizontally, while $j$ runs vertically. 
E.g., quadrant II shows the effect of conditioning right
hemispheric channels on the channels from left hemisphere.

\vspace{1cm}

Fig. 3: Example for a 20x20 $S$-matrix of a 10 sec. segment recorded during
  the seizure free interval using 10 depth electrodes on each side of the 
brain.

\vspace{1cm}

Fig. 4: (A) Average activity pattern in an $8\times 8$ grid electrode;
(B) average passivity and (C) normalized sum of both.

\vspace{1cm}

Fig. 5: Sequence of interdependence patterns $S_{ij}$ 
including preictal (1-14), ictal (15-16) and postictal (17-20) brain electrical activity.

\newpage

\begin{figure}[ht]
\centering
\epsfig{file=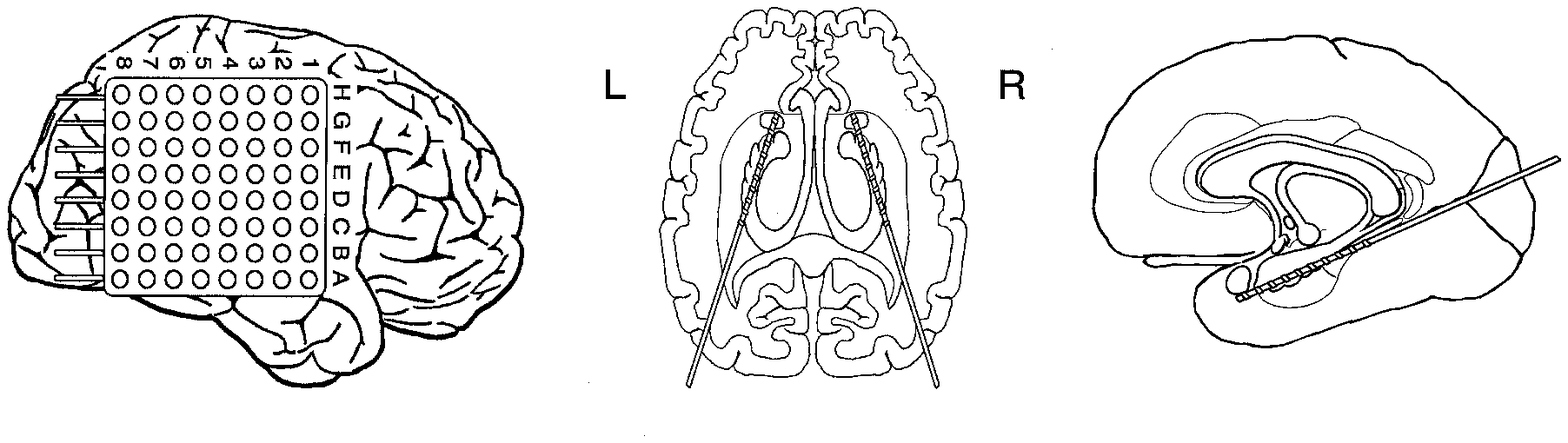,height=4cm,width=13cm}
\vspace{8cm}
\end{figure}
Figure 1\\
Arnhold et al.

\newpage

\begin{figure}[ht]
\centering
\epsfig{file=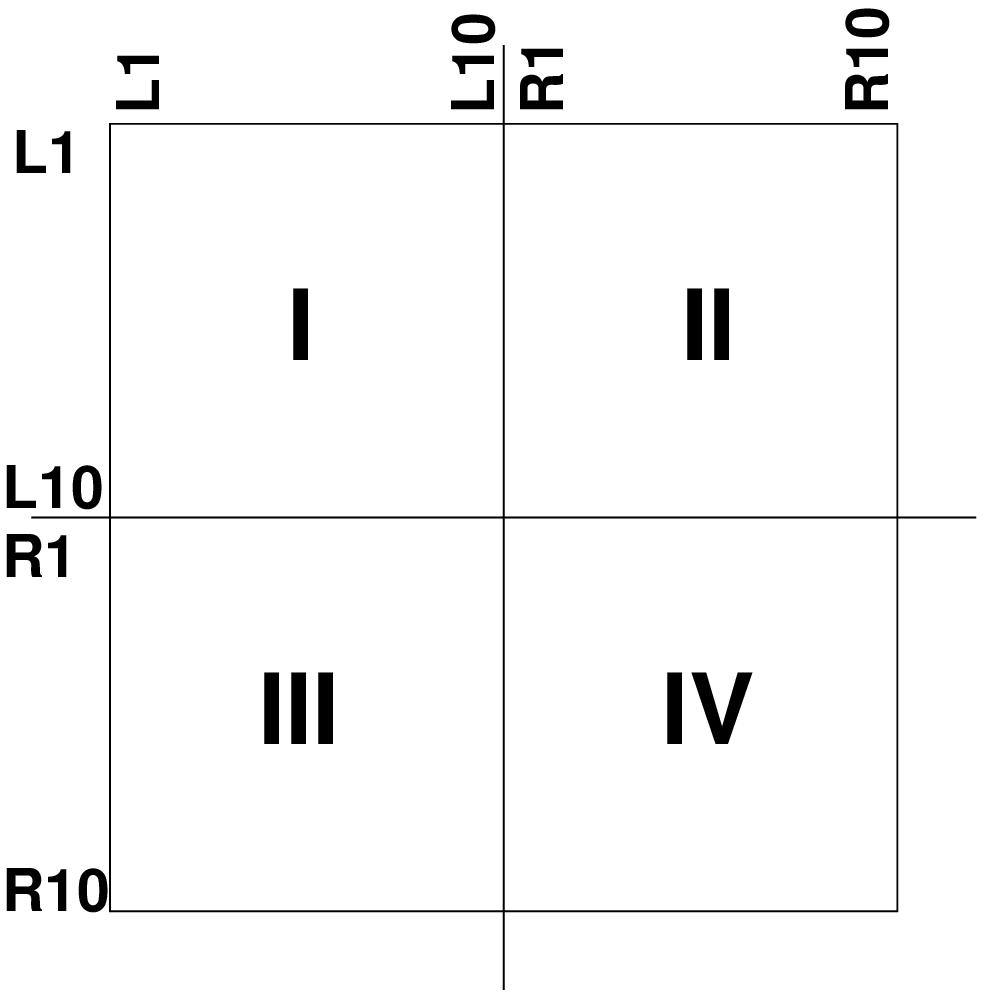,height=8cm,width=8cm}
\vspace{8cm}
\end{figure}
Figure 2\\
Arnhold et al.

\newpage

\begin{figure}[ht]
\centering
\epsfig{file=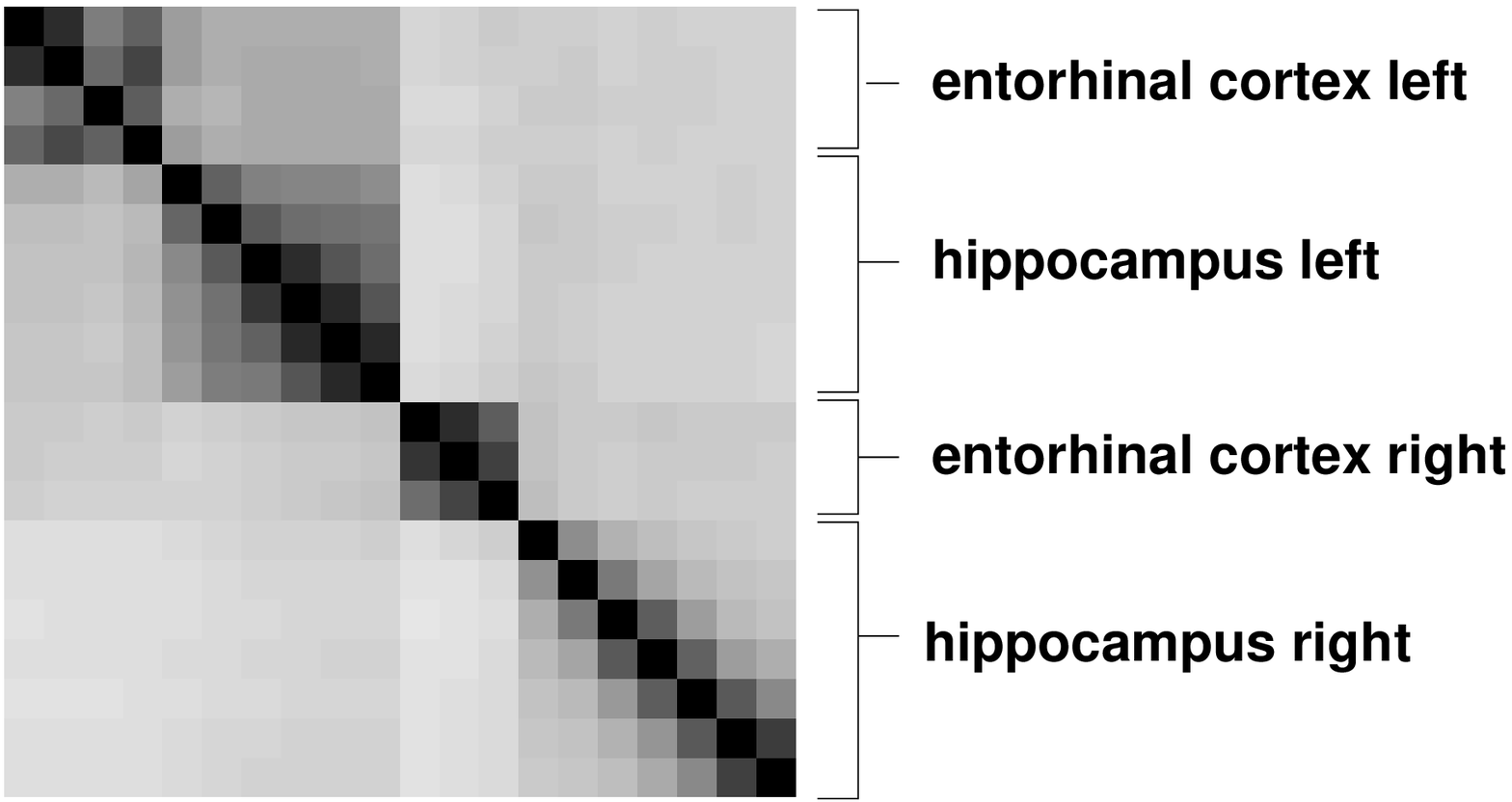,height=7cm,width=13cm}
\end{figure}
\vspace{8cm}
Figure 3\\
Arnhold et al.

\newpage

\begin{figure}[ht]
\centering
\epsfig{file=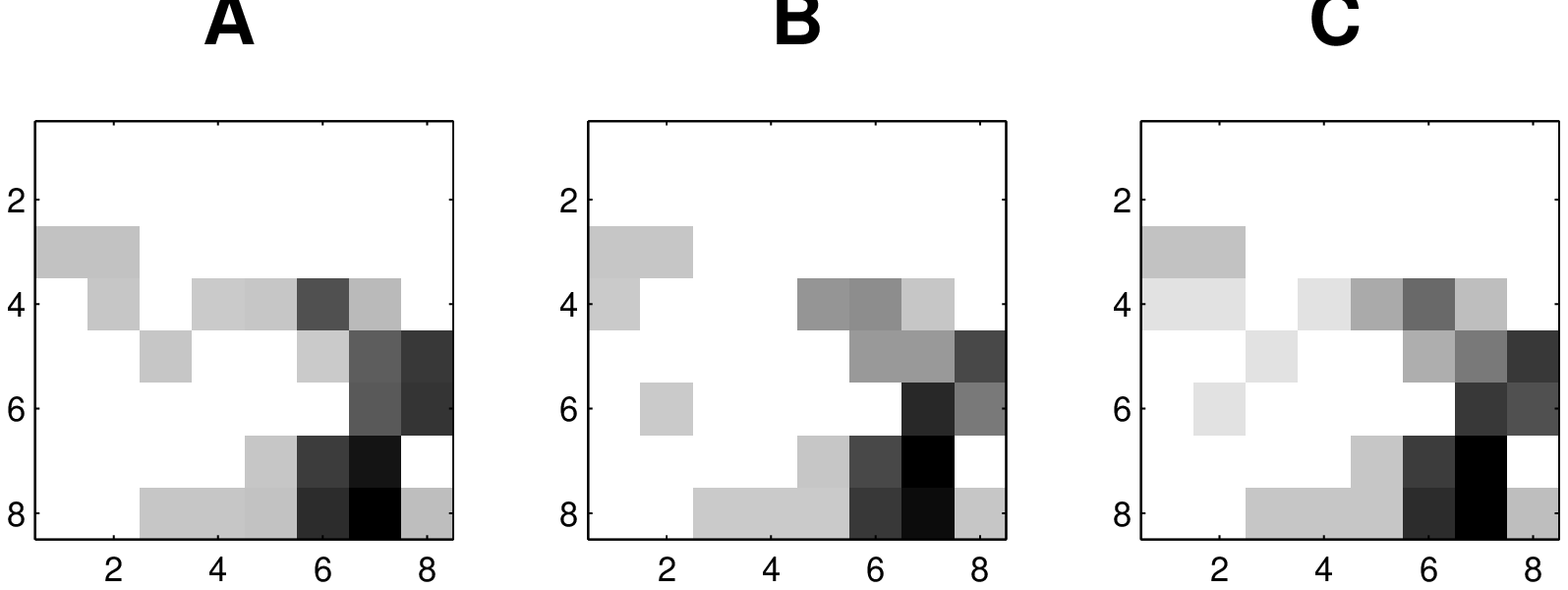,height=14cm,width=5cm,angle=-90}
\end{figure}
\vspace{8cm}
Figure 4\\
Arnhold et al.

\newpage

\begin{figure}[ht]
\centering
\epsfig{file=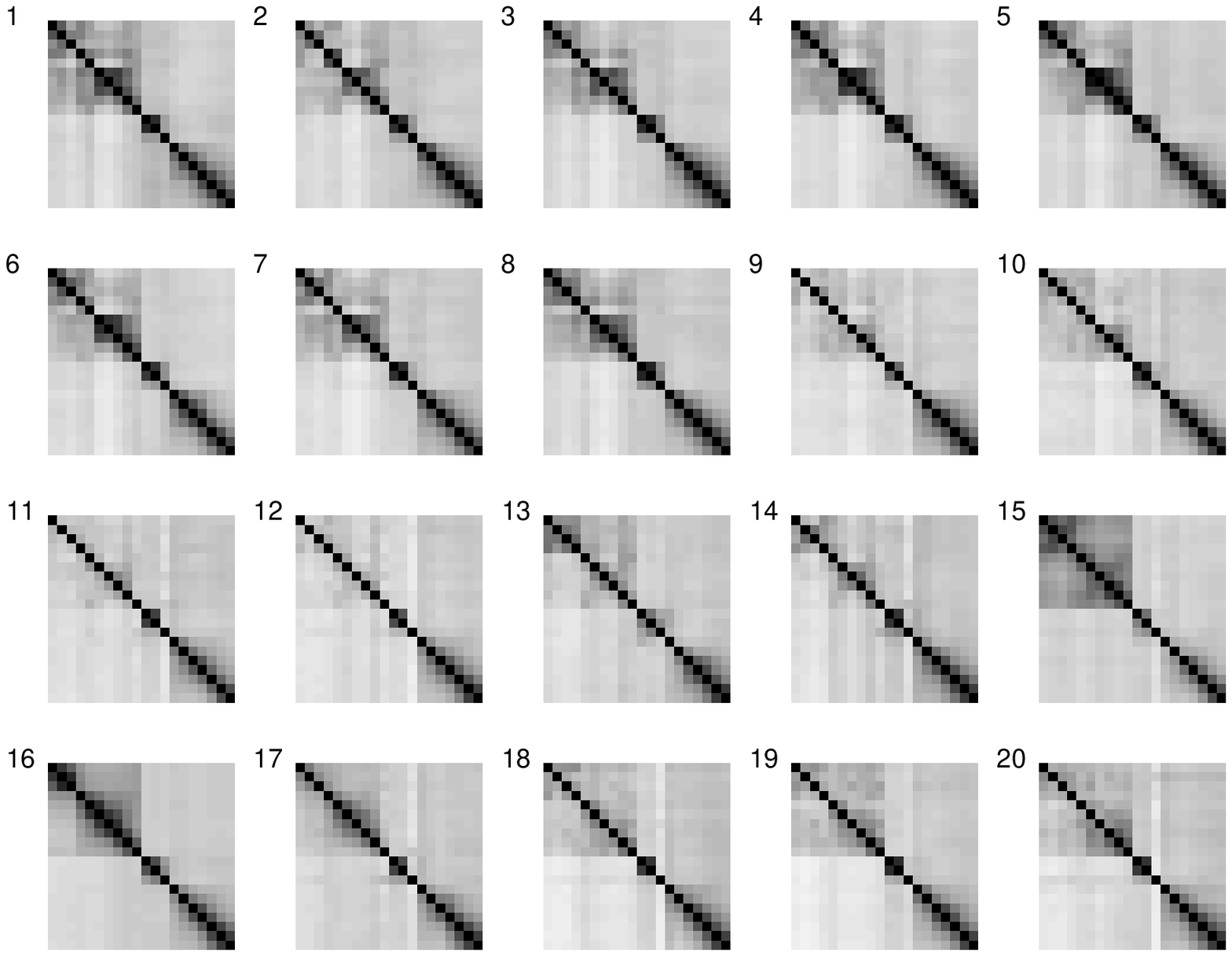,width=14.0cm}
\end{figure}
\vspace{8cm}
Figure 5\\
Arnhold et al.

\end{document}